# The Role of Flexoelectric Coupling and Chemical Strains in the Emergence of Polar Chiral Nano-Structures

Anna N. Morozovska[1*], Salia Cherifi-Hertel[2], Eugene A. Eliseev[3], Victoria V. Khist[4], Riccardo Hertel[2†], and Dean R. Evans[5‡]

[1] Institute of Physics, National Academy of Sciences of Ukraine,
46, pr. Nauky, 03028 Kyiv, Ukraine

[2] Université de Strasbourg, CNRS, Institut de Physique et Chimie des Matériaux de Strasbourg, UMR 7504, 67000 Strasbourg, France

[3] Frantsevich Institute for Problems in Materials Science, National Academy of Sciences of Ukraine
Omeliana Pritsaka str., 3, Kyiv, 03142, Ukraine

[4] Igor Sikorsky Kyiv Polytechnic Institute, Kyiv, Ukraine

[5] Air Force Research Laboratory, Materials and Manufacturing Directorate, Wright-Patterson Air Force Base, Ohio, 45433, USA

**Abstract**

This review examines the conditions that lead to the formation of flexo-sensitive chiral polar structures in thin films and core-shell ferroelectric nanoparticles. It also analyzes possible mechanisms by which the flexoelectric effect impacts the polarization structure in core-shell ferroelectric nanoparticles. Special attention is given to the role of the anisotropic flexoelectric effect in forming a unique type of polarization states with distinct chiral properties, referred to as "flexons".

In the first part of the review, we study the influence of the flexoelectric coupling on the polarity, chirality and branching of metastable labyrinthine domain structures in uniaxial ferroelectric core-shell nanoparticles. We reveal that the transition from sinuous branched domain stripes to spiral-like domains occurs gradually as the flexoelectric coupling strength is increased. Our findings indicate that the joint action of flexoelectric effect and chemical strains, termed as "flexo-chemical" coupling, can significantly influence the effective Curie temperature, polarization distribution, domain morphology, and chirality in multiaxial ferroelectric core-shell nanoparticles. Furthermore, we demonstrate that the combination of flexo-chemical coupling and screening effects leads to the appearance and stabilization of a chiral polarization morphology in nanoflakes of van der Waals ferrielectrics covered by a shell of ionic-electronic screening charge.

---

[*] corresponding author, e-mail: anna.n.morozovska@gmail.com
[†] corresponding author, e-mail: riccardo.hertel@ipcms.unistra.fr
[‡] corresponding author, e-mail: dean.evans@afrl.af.mil

In the second part of the review, we discuss several advanced applications of flexo-sensitive chiral polar structures in core-shell ferroelectric nanoparticles for nanoelectronics elements and cryptography. We underline the possibilities of the flexoelectric control of multiple-degenerated labyrinthine states, which may correspond to a differential negative capacitance (NC) state stabilized in the uniaxial ferroelectric core by the presence of a screening shell. We show that the paraelectric-like state of van der Waals ferrielectric nanoflakes covered by a shell of ionic-electronic screening charge exhibits a pronounced NC effect over a relatively wide range of nanoflake thicknesses, flexo-chemical strains, and surface charge densities.

## 1. Chiral polarization structures induced by the flexoelectric effect

Despite the basic principles for a theoretical description of macro-, meso-, and nanoscale ferroelectrics being established decades ago [1, 2, 3], physical mechanisms of their exclusive polar properties [4, 5, 6] and topological states [7, 8] remain a challenge for fundamental research. These properties, however, hold significant potential across a broad spectrum of advanced applications, including innovative information storage methods that span from low frequencies to THz and optical frequency ranges [9, 10]. Innovative information storage concepts in ferroelectrics rely on the formation of various stable domain structures. These structures may involve a nanoscale modulation of their electric polarization which remains stable in the absence of an applied electric field [11]. A localized electric field above the coercive (or critical) value is required to switch the polarization between the stable (or metastable) states, thus allowing for the domain structure control at the nanoscale, while a non-destructive visualization of domain structures necessitates electric fields significantly lower than the coercive value [12].

Theoretical predictions and experiments have led to the discovery of a great variety of topologically different chiral polarization structures, such as nanoscale polarization vortices [13, 14, 15, 16], skyrmions [17, 18, 19, 20], flux closure domains [21, 22, 23], domain bubbles [24], labyrinthine domains (looking like "branched sinuous" or irregular "mazes") [25, 26], and meandering domain structures [27, 28]. It is quite common for topologically complex chiral polarization structures to originate from the curling of non-Ising type chiral domain walls [29, 30], which can emerge even in classical uniaxial ferroelectrics with nominally 180° ferroelectric domain walls, such as $LiNbO_3$ [31, 32]. In this case, 180° ferroelectric domain walls in $LiNbO_3$ can exhibit complex non-Ising type polarization configurations leading to unique physical properties [29], which offer solid grounds for tailoring nanoscale polar states [32]. Correlations between the three-dimensional (3D) domain wall curvature and its internal polar structure in the form of modulations of the Néel-like character is attributed to the flexoelectric effect [29, 32].

The direct flexoelectric effect describes the appearance of the electric polarization $P_l$ in response to the strain (or stress) gradient $\frac{\partial u_{ij}}{\partial x_k}$ (or $\frac{\partial \sigma_{ij}}{\partial x_k}$). The inverse flexoelectric effect describes the appearance of the strain $u_{ij}$ (or stress $\sigma_{ij}$) in response to the electric polarization gradient $\frac{\partial P_l}{\partial x_k}$ [33, 34]. The flexoelectric coupling is linear; its contribution to the ferroelectric free energy is described by the Lifshitz invariant, $\frac{f_{ijkl}}{2}\left(P_l \frac{\partial u_{ij}}{\partial x_k} - u_{ij}\frac{\partial P_l}{\partial x_k}\right)$, where $f_{ijkl}$ is the flexoelectric effect tensor. The flexoelectric coupling is "omnipresent", because $f_{ijkl}$ has nonzero components for all symmetry groups in solids.

The flexoelectric effect significantly changes the polarization structure in uniaxial ferroelectric thin films with strain gradients [35, 36, 37, 38] and enables the mechanical writing [39] and reading [40] of ferroelectric polarization. This effect is evident in multiaxial ferroelectrics, such as $BaTiO_3$, where the flexoelectric coupling can induce strongly anisotropic Bloch-like [41] and Néel-like [42] domain walls. These structures exhibit chiral and/or achiral characteristics, distinguishing them qualitatively from the classical Bloch-type domain wall structure in ferromagnets.

The synergy of flexoelectric coupling and striction related with the rotation of oxygen octahedra in perovskite ferroelastics (named "roto-striction") can lead to the appearance of inhomogeneous electric fields (named "flexo-roto" field), which are proportional to the structural antiferrodistortive order parameter (oxygen tilt) gradient in perovskite ferroelastics [43]. The proportionality coefficients are the convolution of the flexocoupling and roto-striction tensors. The flexo-roto field can then exist in a wide class of materials with oxygen octahedra rotations [44], such as $SrTiO_3$ [45] and $CaTiO_3$ [46]. The flexo-roto field can contribute significantly to the improper ferroelectricity [47] induced by octahedral rotations at the antiphase boundaries and/or twin walls in $YMnO_3$ [48], $Ca_3Mn_2O_7$ [49] and $CaTiO_3$ [50, 51]. The flexo-antiferrodistortive coupling can lead to the appearance of versatile spatially modulated structures in multiferroics [52, 53].

The curling of non-Ising domain walls can lead to the formation of helicoidal domain patterns, such as vortices and skyrmions. Possible mechanisms and driving forces of these formations, which lead to helicoidal patterns in meso- and nanoscale ferroelectrics, are not fully understood. The primary mechanism may be flexoelectricity, which strongly affects the minimum of the electrostatic energy of internal (depolarization and/or stray) electric fields in the presence of curved surfaces, elastic defects, and/or charged defects [54, 55, 56]. Another possible mechanism is the stabilization of the helicoidal domain patterns by an antisymmetric coupling akin to the Dzyaloshinskii-Moryia interaction (DMI) free energy term, which plays a central role in certain categories of ferromagnets [57, 58] and which is predicted to exist also in ferroelectric materials due to the inversion symmetry breaking [59]. The occurrence of such an energy term is also predicted by first-principles simulation results [60].

The thermodynamic description of the flexoelectric effect involves the Lifshitz invariant in the elastic contribution $G_{es+flexo}$ to the Landau-Ginzburg-Devonshire (LGD) free energy [61, 62]:

$$G_{es+flexo} = \int_V d^3r \left( -\frac{s_{ijkl}}{2}\sigma_{ij}\sigma_{kl} - Q_{ijkl}\sigma_{ij}P_kP_l - \frac{F_{ijkl}}{2}\left(\sigma_{ij}\frac{\partial P_l}{\partial x_k} - P_l\frac{\partial \sigma_{ij}}{\partial x_k}\right)\right). \qquad (1)$$

Here $V$ is the volume of a ferroelectric, $\sigma_{ij}$ is the stress tensor, $s_{ijkl}$ are elastic compliances, $Q_{ijkl}$ and $F_{ijkl}$ are the electrostriction and flexoelectric tensor components, respectively, and $P_i$ are the polarization components. The subscripts $i, j, k$, and $l$ are Cartesian indices with values of 1, 2, and 3. Finite element modeling (FEM) uses the LGD free energy for simulation of the polarization distribution in ferroelectrics.

The ferroelectric analogue of the DMI was considered by Strukov and Levanyuk [63] in the context of Lifshitz invariants. It was shown that the flexoelectric coupling may have an influence on the formation and morphology of polar chiral structures in ferroelectrics similar to that of the DMI in ferromagnets [64]. It was shown therein [64] that an anisotropic flexoelectric effect can give rise to a specific type of polarization state, termed "flexon", with distinct chiral properties. Flexons are ferroelectric polar textures in a cylindrical core-shell nanoparticle, which reveal a chiral polarization structure containing two oppositely oriented diffuse axial domains located near the cylinder ends, separated by a region with a zero-axial polarization (see the top row in **Fig. 1**). The flexon chirality can be switched by reversing the sign of the corresponding flexoelectric coefficient; furthermore, the anisotropy of the flexoelectric coupling tensor influences electric polarization and domain morphology in the nanoparticle in a principal way. In the azimuthal plane, the flexon displays the polarization state of a vortex with an axially polarized core region, i.e., a structure with the topological hallmark of a meron (see the bottom row in **Fig. 1**). The flexon has a rounded shape and a flexo-sensitive chirality, being localized near the nanoparticle surface, it is reminiscent of "chiral bobber" structures in magnetism [65]. Since the flexoelectric effect is responsible for the stabilization of chiral flexons, the similarities between the flexoelectric interaction and the ferroelectric DMI is evident in this case [64]. The relatively wide temperature range (from 200 to 400 K) where flexons exist in BaTiO$_3$ nanocylinders gives hope that they can be observed experimentally by scanning-probe piezoresponse force microscopy and nonlinear optical microscopy methods.

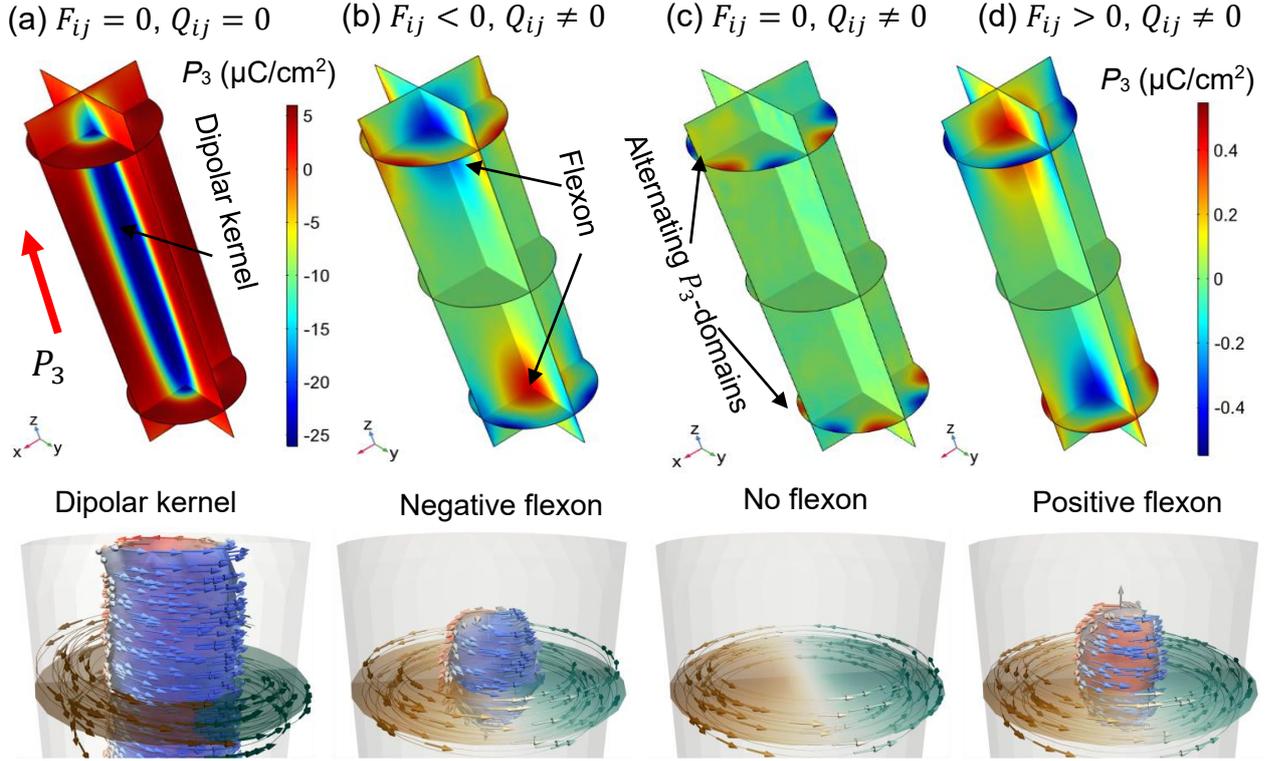

**FIGURE 1. Chiral polarization textures, induced by the flexoelectric effect in ferroelectric nanocylinders.** The top row: a distribution of the polarization component $P_3$ inside a cylindrical BaTiO$_3$ nanoparticle of radius $R$, covered with an elastically soft semiconducting shell with a thickness $\Delta R \ll R$, placed in an isotropic elastically soft effective medium. The distributions are calculated: **(a)** without electrostriction coupling ($Q_{ij} = 0$) and flexoelectric effect ($F_{ij} = 0$); and with electrostriction coupling ($Q_{ij} \neq 0$) and **(b)** negative, **(c)** zero, and **(d)** positive values of flexoelectric coefficients $F_{ij}$. The left color scale corresponds to the $P_3$ distribution in **(a)**, and the right color scale corresponds to the $P_3$ distributions in **(a)-(d).** The bottom row: a magnified view of the flexon structure, where the arrows show the orientation of the polarization vector $\vec{P}$. Adapted from Ref. [64].

The appearance of flexo-sensitive polarization vortices is possible in thin ferroelectric films under certain conditions (see Ref. [66] and references therein), when a change of the flexoelectric coefficient sign leads to a reorientation of the vortex core axial polarization, making the flexo-sensitive 3D vortices similar to the flexons in cylindrical nanoparticles [64]. The spike-like cores of the 3D vortices with in-plane vorticity are located near the film – dead-layer interfaces (see **Fig. 2(a)**). The flexoelectric coupling induces an out-of-plane quadrupolar moment of the cores. The flexo-sensitive vortices generate vortex-antivortex pairs, with the antivortices exhibiting in-plane anti-circulation. These antivortices have smooth, wide dipolar cores that extend throughout the film, and their shape and other characteristics remain largely unaffected by the coupling. The topology of the ferroelectric vortex and antivortex can be distinguished by the winding number $W$ [67], where $W = +1$ for the vortex and $W = -1$ for the antivortex.

Without flexoelectric coupling (i.e., $F_{ij} = 0$), the polarization component $P_y$ is nonzero due to the electrostriction coupling, and two Bloch points, where $\vec{P} = 0$, develop symmetrically under the top and bottom surfaces of the ferroelectric film (see dotted curve in **Fig. 2(d)**). The depth profile of $P_y$ becomes asymmetric with respect to the film surfaces for $F_{ij} \neq 0$, and a single Bloch point exists in this case (see **Fig. 2(e)**). The transformation $F_{ij} \rightarrow -F_{ij}$ leads to $P_y \rightarrow -P_y$ (compare red and blue curves in **Fig. 2(e)**). Hence, the sign of $P_y$ is defined by the sign of $F_{ij}$, since the flexo-induced contribution to the polarization component $P_y$ vastly dominates over the electrostriction contribution. These results demonstrate the flexon-type character [64] of the vortex polarization, which is also in agreement with the fact that a Bloch point carries a topological charge of magnitude 1; this can be corroborated by the calculation of a corresponding topological index, which quantifies the chirality of the polarization structure. The topological index [68]:

$$n(y) = \frac{1}{4\pi} \int_S \vec{p} \left[ \frac{\partial \vec{p}}{\partial x} \times \frac{\partial \vec{p}}{\partial z} \right] dx dz, \qquad (2)$$

where $\vec{p} = \frac{\vec{P}}{P}$ is the unit vector of the polarization orientation. The integration in Eq.(2) is performed over the vortex cross-section $S = \{x, z\}$. The dependence $n(y)$ is shown in **Fig. 2(f)** for zero, positive, and negative $F_{ij}$. In all cases, $n(y) = \pm 1/2$ at the film surfaces. Since the polarization magnitude $P(0, y, 0)$ coincides with $|P_y(0, y, 0)|$ at the vortex axis, and $P_y(0, y, 0) = P(0, y, 0) = 0$ in the Bloch point, the topological index jumps at that point from -½ to +½ (see the red curve in **Fig. 2(f)**), or from +½ to -½ (see the blue curve in **Fig. 2(e)**), depending on the $F_{ij}$ sign. The topological index, which can be interpreted as the degree to which a structure is chiral, changes its sign from one surface to the other, and changes the sign upon reversal of the sign of $F_{ij}$. These observations provide a clear correlation between the flexoelectric coupling and the formation of chiral polar vortices in thin ferroelectric films. The surface localization of the flexo-sensitive vortices makes them similar to the "edge" localization of flexons in cylindrical nanoparticles [64].

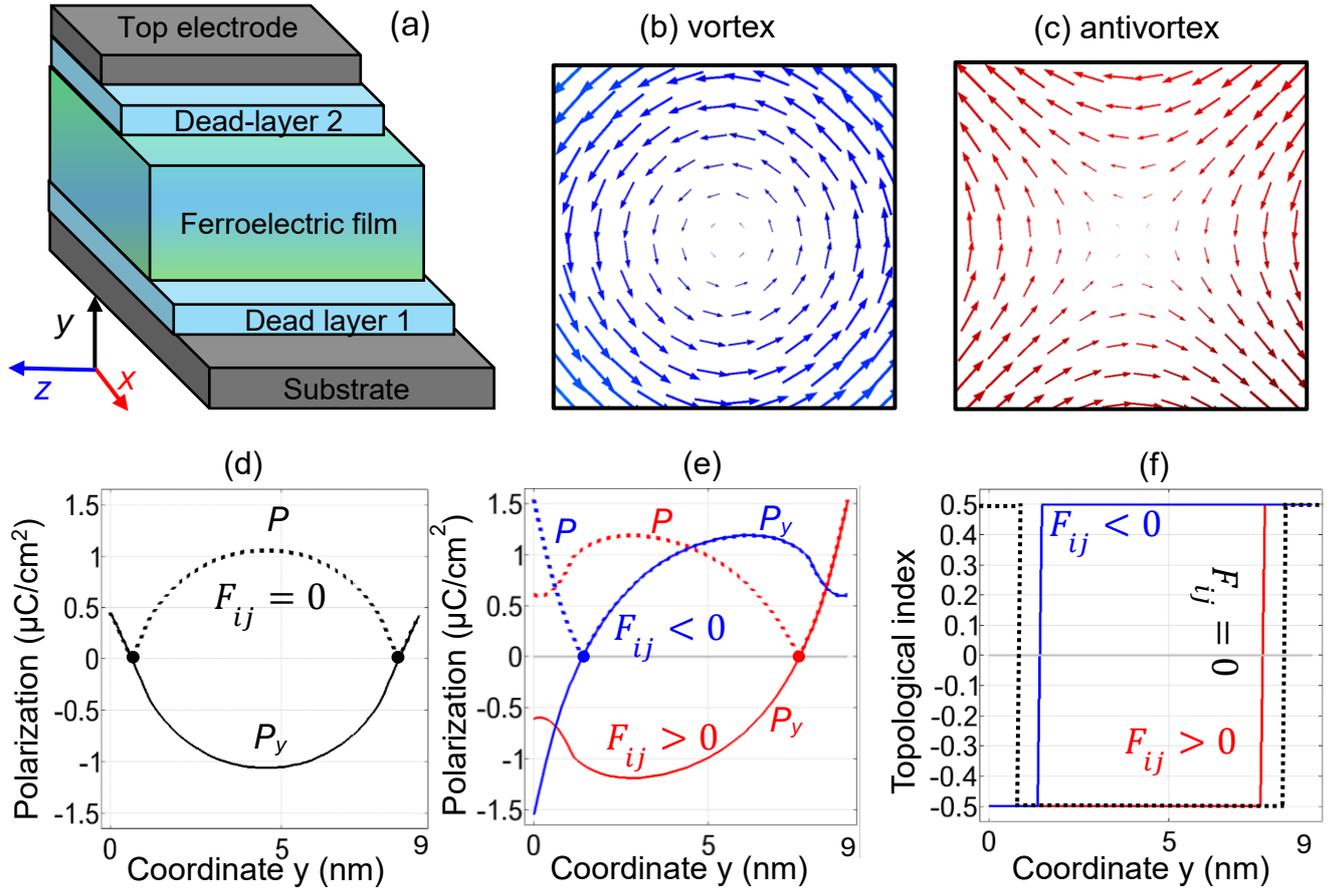

**FIGURE 2. Flexo-sensitive polarization vortices in ferroelectric thin films. (a)** The ferroelectric film is placed between two paraelectric dead layers with very high dielectric permittivity. The coordinate frame is shown in the bottom left corner. A ferroelectric vortex **(b)** and antivortex **(c)** are shown, whose topology can be distinguished by the winding number $W$. **(d-e)** Depth profiles of the out-of-plane polarization component $P_y$ (solid curves) and magnitude $P$ (dotted curves) calculated at the vortex axis for zero (black curves, d), negative (blue curves, e) and positive (red curves, e) flexoelectric coefficients $F_{ij}$. The thick circular dots represent Bloch points, where $P = 0$. **(f)** The y-profile of the polarization's topological index $n(y)$ is plotted for zero (black dotted lines), positive (solid red lines), and negative (solid blue lines) $F_{ij}$. Figures **(d)-(f)** are calculated for a 9-nm thick BaTiO$_3$ film, mismatch strain $u_m = 0.1\%$, and $T = 300$ K. Adapted from Ref. [66].

The main focus of research on the chiral structures in nanoscale ferromagnets and ferroelectrics has been centered around thin films and multilayers (see e.g., topical reviews [7, 8, 69]). The growing interest in 0D ferroelectricity [6] has spurred research on polar chiral structures in ferroelectric nanoparticles, particularly vortex-like [70, 71] and labyrinthine [72, 73, 74] configurations. This research is further motivated by the ongoing efforts to synthesize Si-compatible and inexpensive ferroelectric nanoparticles [75, 76, 77], which hold promises for various applications. Moreover, the influence of metastable domain configurations on negative capacitance states in these nanoparticles [78, 79] has attracted increasing attention from both fundamental and applied perspectives. Notably, the flexoelectric effect [71, 80] and chemical strains [74, 81, 82] caused by topological and/or point defects (such as bulk and/or surface vacancies and ions) can lead to principal changes in the polar states and domain structure morphology.

Without the flexoelectric coupling, three-dimensional vortex states with a kernel, which are found in BaTiO3 core-shell nanoparticles, exhibit a manifold degeneracy. This degeneracy results from three factors: (1) three equiprobable directions of the vortex axis, (2) clockwise and counterclockwise polarization rotation along the vortex axis, and (3) two polarization directions in the kernel [70]. This manifold degeneracy of the vortex states in a single nanoparticle can be utilized for applications as multi-bit memory elements and associated logic units. The free rotation of a vortex kernel, possible for nanoparticles in a soft matter medium with a controllable viscosity, may be used to imitate qubit features. However, a single polarization vortex with a dipolar kernel can be stable if the electrostriction coupling is relatively weak in the nanoparticle core covered with an elastically soft shell [71]. For a core covered with a rigid shell, the anisotropic elastic properties of the shell can stabilize vortex-like structures with three flux-closure domains [71]. A mismatch strain between the core and the shell compensates the curling of the flux-closure domains in the core confined by an elastically-anisotropic rigid shell. An illustration of the ferroelectric core covered by a soft or rigid shell is shown in **Fig. 3(a).**

In contrast to the case without flexoelectric coupling, the introduction of flexoelectricity significantly alters domain wall morphology. The flexoelectric coupling leads to a noticeable curling of the flux-closure domain structure in the BaTiO3 core-shell nanoparticles. Moreover, the flexoelectric coupling influences topological defects such as Bloch point structures and Ising lines, where $\vec{P} = 0$ (see Ref. [83]); they are shown in **Fig. 3(c)-(f)**. The purple spheres indicate the position of Bloch points, determined by the intersection points of the three iso-surfaces $P_1 = 0$, $P_2 = 0$, and $P_3 = 0$.

The topological structure of polarization inside the core covered with a soft shell, calculated without flexoelectric coupling ($F_{ij} = 0$) and isotropic electrostriction ($Q_{11}^{c,s} \approx Q_{12}^{c,s}$), does not contain Bloch point structures (see **Fig. 3(b)**). Anisotropic electrostriction ($Q_{11}^{c,s} \neq Q_{12}^{c,s}$) leads to the appearance of a one-dimensional topological line defect with $|\boldsymbol{P}| = 0$ known as "Ising line" (see **Fig. 3(c)**). The flexoelectric coupling ($F_{ij} \neq 0$) transforms the Ising line into the two Bloch points located at opposite sides of the domain wall (see **Fig. 3(d)**). Although these Bloch points are very close to the surface, they are not located at diametrically opposite positions [71].

The topological structure of polarization, calculated without flexoelectric coupling ($F_{ij} = 0$) and mismatch strain ($u_m = 0$), contains an Ising line (see **Fig. 3(e)**). A flexoelectric coupling ($F_{ij} \neq 0$) transforms the Ising line into a single Bloch point located in the core center (see **Fig. 3(f)**).

Introducing a tensile or compressive mismatch strain ($u_m \neq 0$) does not change the polarization topology and the position of the central Bloch point (see **Fig. 3(g)**).

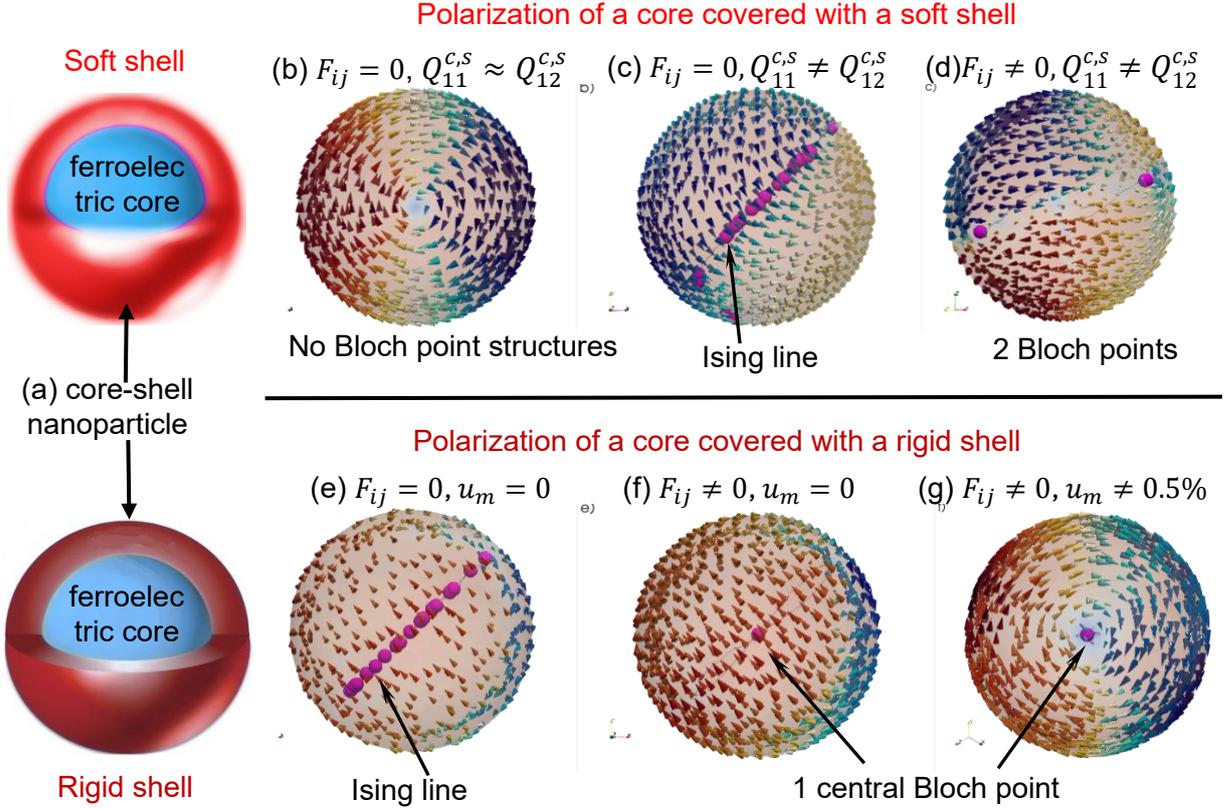

FIGURE 3. **Flexo-elastic control factors of domain morphology in core-shell ferroelectric nanoparticles. (a)** An illustration of a ferroelectric core covered by a soft or rigid shell**.** Bloch Point morphologies in a ferroelectric core covered with a soft (panels **(b-d)**) or a rigid (panels **(e-g)**) shell. The flexoelectric effect is absent ($F_{ij} = 0$) for panels **(b, c, e)** and present ($F_{ij} \neq 0$) for panels **(d, f, g)**. The electrostriction anisotropy is small ($Q_{11}^{c,s} \approx Q_{12}^{c,s}$) for panel **(b)** and high ($Q_{11}^{c,s} \neq Q_{12}^{c,s}$) for panels **(c-g)**. A mismatch strain is absent ($u_m = 0$) for panels **(b, c, e)** and present ($u_m \neq 0$) for panels **(d, f, g)**. The radius of BaTiO$_3$ core $R = 10$ nm, shell thickness $\Delta R = 4$ nm, and $T = 298$ K. Anisotropic electrostriction and flexoelectric coefficients correspond to BaTiO$_3$. Adapted from Ref. [71].

The flexoelectric coupling can control the polarity and chirality of equilibrium polar domain structures in uniaxial ferroelectric core-shell nanoparticles [64, 73]. Under certain screening conditions at the particle surface and for a definite range of particle sizes, stable labyrinths evolve spontaneously from a random initial polarization distribution with arbitrarily small domains. While both mazes and spirals share a fundamental spiral-like arrangement, they differ significantly in their characteristic size and regularity. Mazes exhibit a fine-scale spiral structure with less pronounced definition, which can be viewed as a precursor to the larger, more pronounced spiral domains that emerge with increasing flexoelectric coupling. The effects of flexoelectric coupling may significantly decrease the number of "branching points" in the maze, increase its scale (i.e., make the sinuous-like domains fatter) and/or invert its chirality. However, it does not fundamentally change the overall shape of the maze or the system general response to the applied field.

To illustrate the role of the flexoelectric coupling, we apply a bipolar voltage pulse, with the time dependence depicted in **Fig. 4(a)**, to the spherical core-shell nanoparticle. We then simulate the electric polarization distribution inside the core composed of Sn$_2$P$_2$S$_6$, which can be treated as a uniaxial ferroelectric with relatively high accuracy for small nanoparticles [73]. Initial state and relaxed polarization distributions calculated for zero ($F_{ij} = 0$) and high positive flexoelectric coefficients $F_{ij}$ at different moments of time (marked in **Fig. 4(a)** by the numbers from 0 to 5) are shown in **Fig. 4(b)** and **4(c)**, respectively. The characteristic time of polarization distribution relaxation is 1-2 orders of magnitude longer than the Khalatnikov time (constant $\tau \sim 10^{-10}$s), we use a simulation time of at least $10^3 \tau$. This ensures that the polarization distribution stabilizes well before the simulation ends.

For a definite range of particle radii (approximately from 5 nm to 20 nm), effective screening lengths at the particle surface (approximately from 1 pm to 0.1 nm, and temperatures (approximately from 270 K to 310 K), stable branched sinuous stripes and/or chiral spiral-like domain structures evolve spontaneously from a random initial polarization distribution with arbitrarily small domains and gradually disappear under a voltage increase (see the image sequence of the polarization distributions in the equatorial cross-section $\{X_1, X_2\}$ of the core-shell nanoparticle in **Fig. 4(b)** and **4(c)**).

Without flexoelectric coupling, the maze polarity is controlled by the field projection on the particle polar axis $X_3$ at voltages higher than the critical value [64], while voltages below the critical value do not have an influence on the mazes in any noticeable way in comparison with the off-field relaxation. The maze polarity controlled by the field projection is illustrated in the bottom images "3" and "5" in **Fig. 4(b)**. In these images, the color contrast is reversed in the mazes that have relaxed after the maximal voltage +1.5 V or -1.5V is switched off. Notably, mazes "3" and "5", which form after the application and removal of electric field, strongly resemble the slightly less pronounced maze "1", which forms through relaxation from the initial random polarization state "0". This resemblance, along with the reversed color patterns in mazes "3" and "5", suggests that the system exhibits a memory effect, preferentially returning to a similar maze-like configuration after the field is removed, even when the field polarity is reversed.

While the presence of strong flexoelectric coupling does not alter the general formation of spiral-like domains, it does significantly influence their characteristic size. Similar to the case of no flexoelectric coupling, these structures evolve spontaneously from a random initial polarization distribution and gradually disappear under a voltage increase (see the top row in **Fig. 4(c)**). The polarity of spiral-like domains is controlled by the field projection on the particle polar axis at voltages higher than the critical value, as illustrated by bottom images "3" and "5" in **Fig. 4(c)**, where the color contrast is reversed in the spiral-like domains relaxed after the maximal voltage +1.5 V or -1.5V is switched off. It is important to note that, as was also observed in the case without flexoelectric coupling, the spiral-like domains are irregular and an ideal circle-like domain in the radial cross-section $\{X_1, X_2\}$ of the nanoparticle core is never observed. However, strong flexoelectric coupling does lead to a noticeable increase in the scale of these spiral-like domains, resulting in thicker spiral patterns (with a smaller number of branching points) compared to the finer branched meander-like domain patterns (with a larger number of branching points) observed without the coupling.

The transition from fine branched-like mazes (with breaks and branching points) to larger scale spiral-like mazes is gradual and dependent on the magnitude of $F_{ij}$ as illustrated in **Fig. 5** for the core-shell $Sn_2P_2S_6$ nanoparticle with radius $R = 16$ nm. **Figure 5(a)** represents the case without flexoelectricity ($F_{ij} = 0$), showing a sinuous branch-like maze with several breaks and branching points. As $F_{ij}$ increases to positive values of $F_{11} = 10^{-11} m^3/C$, $F_{12} = 0.9 \cdot 10^{-11} m^3/C$, and $F_{44} = 3 \cdot 10^{-11} m^3/C$ (**Fig. 5(b)**), a transition towards larger-scale spiral-like domains begins. This trend continues as the positive values of $F_{ij}$ are doubled (**Fig. 5(c)**) and tripled (**Fig. 5(d)**), further emphasizing the influence of flexoelectric coupling strength on domain structure. It is seen that branch-like and spiral-like domain structures are observed in the equatorial cross-section $\{X_1, X_2\}$ of the core, while the counter-polarized domain stripes are observed in the polar cross-section $\{X_1, X_3\}$ of the core (compare the top and bottom rows in **Fig. 5**). It is worth noting that the transition from a fine branch-like to a larger-scale spiral-like domain structure in uniaxial core-shell nanoparticles is insensitive to the sign of the flexoelectric coefficients ($F_{ij}$). This transition remains consistent even when the sign of all $F_{ij}$ components is reversed, $F_{ij} \rightarrow -F_{ij}$. While the transition is unaffected by the sign of the flexoelectric tensor, it does exhibit sensitivity to its anisotropy. Further investigation is required to fully understand this aspect.

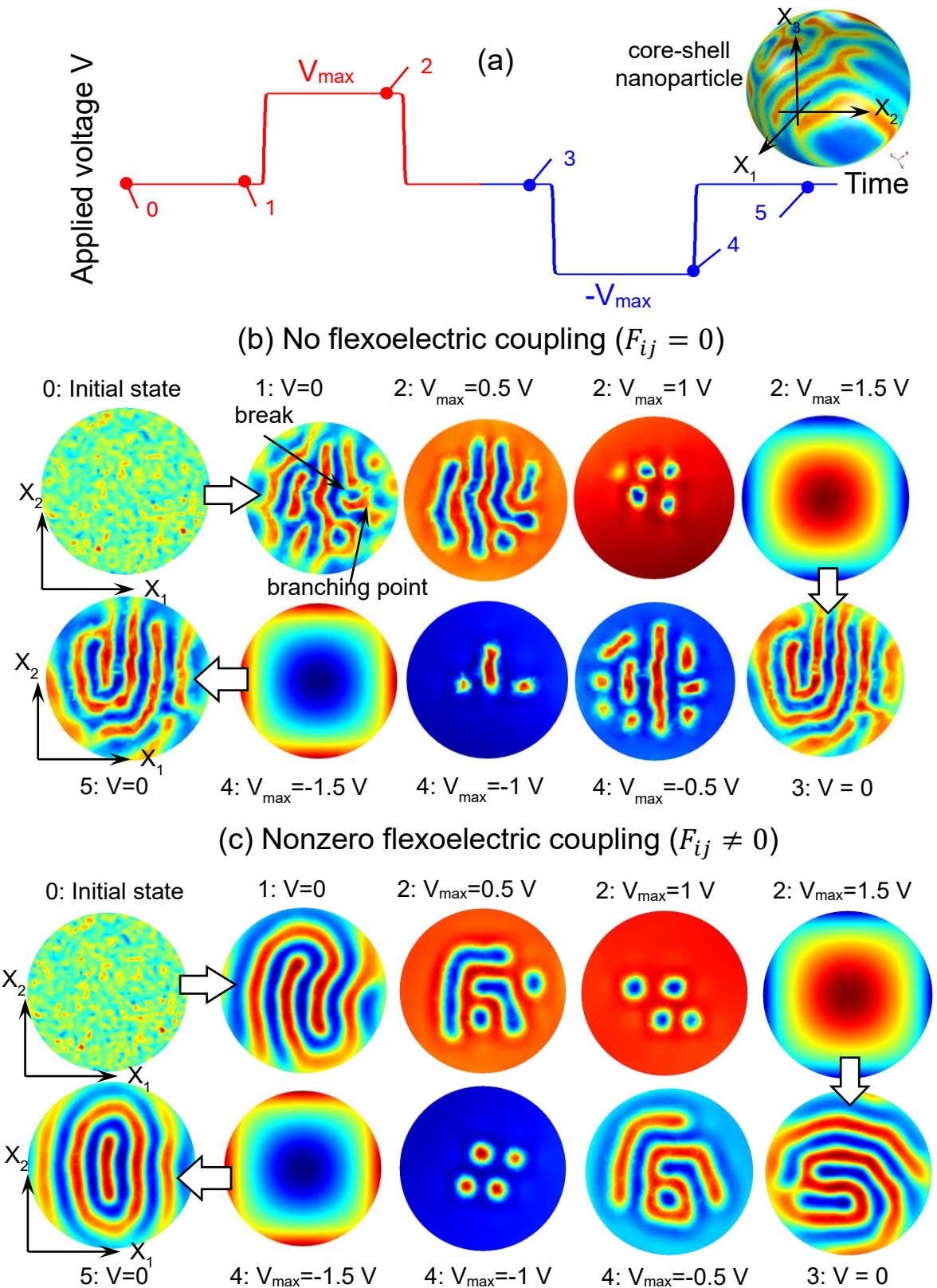

**FIGURE 4. Flexoelectric control of mazes polarity in core-shell ferroelectric nanoparticles.** (a) Shape of applied voltage pulse. Parts **(b)** and **(c)** are polarization distributions calculated for zero and high positive flexoelectric coefficients $F_{ij}$ at different voltages $V$ (listed in the plots). Panels in (b) illustrate mazes, characterized by their fine-scale spiral structure with breaks and branching points, while panels in (c) show the larger, more pronounced spiral domains. The relaxed domain structures "3" and "5" recover spontaneously

after the positive (+1.5 V) or negative (-1.5 V) voltage is switched off. Material parameters correspond to the $Sn_2P_2S_6$ nanoparticles listed in Ref. [64], core radius $R = 16$ nm, screening length $\lambda = 9$ pm, room temperature, and flexoelectric coefficients $F_{ij} = 0$ for part **(b)**, and $F_{11} = 3$, $F_{12} = 2.7$, $F_{44} = 9$ (in $10^{-11}$ m$^3$/C) for part **(c)**. Part **(b)** of the figure is adapted from Ref. [64].

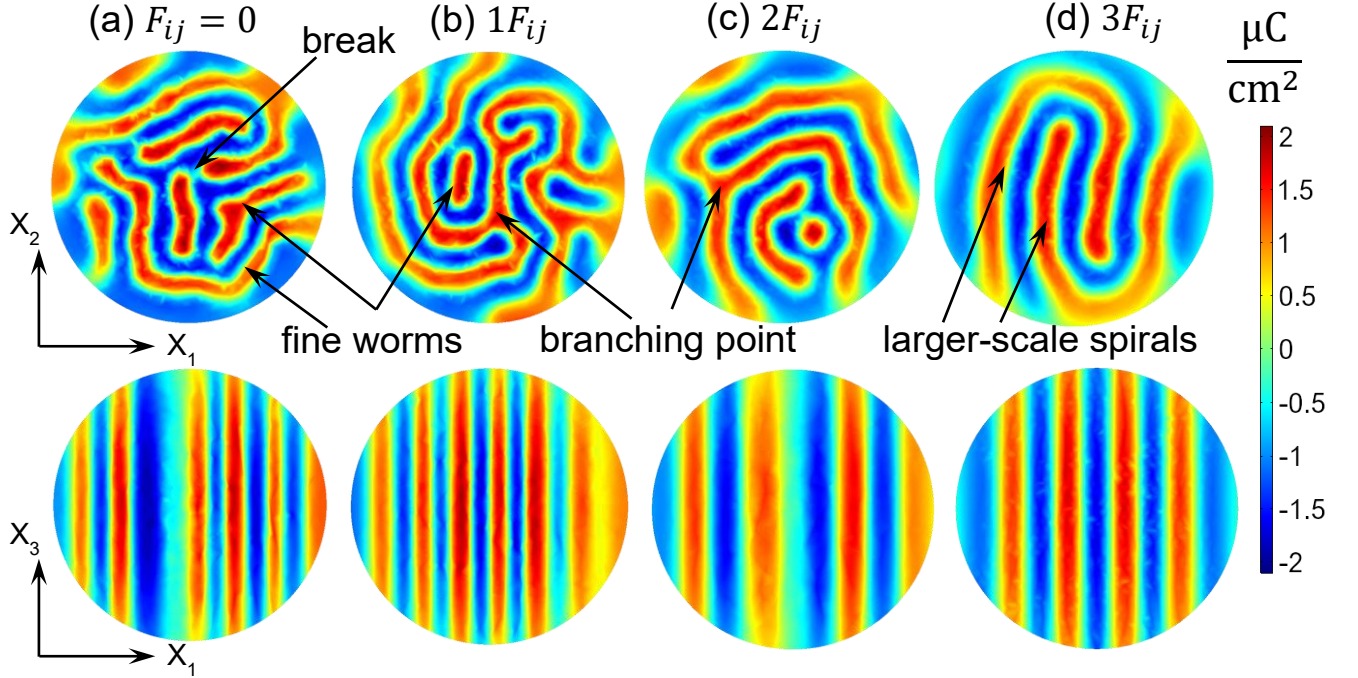

**FIGURE 5**. **The evolution of relaxed domain morphology with increasing flexoelectric coupling in a spherical core-shell Sn$_2$P$_2$S$_6$ nanoparticle.** Top row shows chiral domain structures in the equatorial cross-section $\{X_1, X_2\}$. Bottom row shows counter-polarized domain stripes in the polar cross-section $\{X_1, X_3\}$. Calculations are performed for the core radius $R = 16$ nm, screening length $\lambda = 9$ pm, room temperature, and different values of flexoelectric coefficients: **(a)** $F_{ij} = 0$, **(b)** $F_{11} = 1$, $F_{12} = 0.9$, $F_{44} = 3$ (in $10^{-11}$ m$^3$/C), **(c)** $F_{11} = 2$, $F_{12} = 1.8$, $F_{44} = 6$ (in $10^{-11}$ m$^3$/C), and **(d)** $F_{11} = 3$, $F_{12} = 2.7$, $F_{44} = 9$ (in $10^{-11}$ m$^3$/C). External voltage is absent.

Note that the above-described changes of the domain structure morphology occur in relatively large spherical core-shell Sn$_2$P$_2$S$_6$ nanoparticles ($R = 16$ nm). For comparison, **Fig. 6** illustrates the changes of the domain structure morphology for the core-shell nanoparticle with a radius half that size ($R = 8$ nm). It is seen that a typical maze (with sinuous-like domains and branching points) is stable in the equatorial cross-section $\{X_1, X_2\}$ of the core-shell nanoparticle for $F_{ij} = 0$ (**Fig. 6(a)**). As $F_{ij}$ increases, the maze structure has a similar shape and thicker details (**Fig. 6(b)**). The domain pattern becomes lower-contrast and less branched with a two-fold increase of $F_{ij}$ (**Fig. 6(c)**). The maze eventually transforms into lower-contrast curved stripes under a three-fold increase of $F_{ij}$ (**Fig. 6(d)**). In contrast to the larger nanoparticles (**Fig. 5**), the smaller nanoparticles (**Fig. 6**) exhibit a suppression of the branched-like to spiral-like domain transition with increasing $F_{ij}$, ultimately resulting in lower-contrast curved stripes.

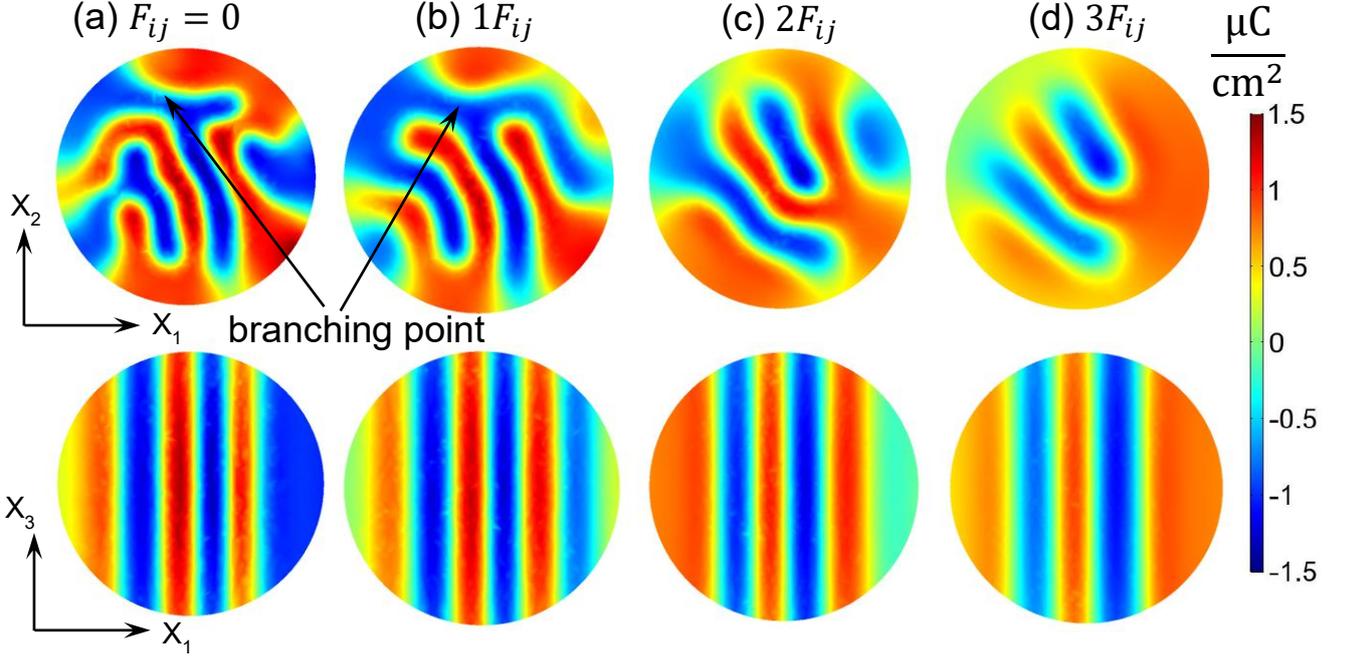

**FIGURE 6**. **The influence of flexoelectric coupling on relaxed polarization distributions in a spherical core-shell $Sn_2P_2S_6$ nanoparticle.** Top row shows chiral domain structures in the equatorial cross-section $\{X_1, X_2\}$. Bottom row shows counter-polarized domain stripes in the polar cross-section $\{X_1, X_3\}$. Calculations are performed for the core radius $R = 8$ nm, screening length $\lambda = 4.4$ pm, room temperature, and different values of flexoelectric coefficients: **(a)** $F_{ij} = 0$, **(b)** $F_{11} = 1$, $F_{12} = 0.9$, $F_{44} = 3$ (in $10^{-11}$m³/C), **(c)** $F_{11} = 2$, $F_{12} = 1.8$, $F_{44} = 6$ (in $10^{-11}$m³/C), and **(d)** $F_{11} = 3$, $F_{12} = 2.7$, $F_{44} = 9$ (in $10^{-11}$m³/C). External voltage is absent.

Prior studies using FEM simulations have demonstrated that an increasing flexoelectric coupling can lead to the disappearance of maze-like domain patterns in smaller, ellipsoidal ferroelectric nanoparticles. This effect, observed in Ref. [54], is caused by an increase in the effective polarization gradient coefficients, where renormalization is proportional to the second power of the flexocoupling coefficients $F_{ij}$. Furthermore, these studies reveal an increase in average domain wall spacing and a decrease in domain contrast within the polar cross-section $\{X_1, X_3\}$ with increasing flexoelectric coupling. In this sense, increasing the flexoelectric coupling evokes a response similar to the gradient-induced morphological phase transition described in Ref. [25], ultimately leading to the suppression of maze patterns. It is crucial to distinguish this suppression from the "geometric catastrophe" effect in ferroelectrics [25, 84], where domain structures vanish due to decreasing particle radius, not increasing flexoelectric coupling.

The observed influence of flexoelectric coupling on domain structures has broader implications beyond the specific cases discussed above. Given the general validity of the LGD approach, the principle of flexoelectric domain control is likely applicable to a wide range of core-shell ferroelectric nanoparticles. This capability to control domain patterns through flexoelectricity opens exciting possibilities for technological applications, particularly in advanced cryptography, which would benefit from precise control over nanoscale domain structures.

## 2. Potential applications of flexo-sensitive chiral polar structures in core-shell ferroelectric nanoparticles

The potential applications of flexo-sensitive chiral polar structures in core-shell ferroelectric nanoparticles are anticipated to be similar to those of ferromagnets (see e.g., Ref. [85] and refs. therein), but with key differences. Both ferromagnetic and ferroelectric chiral polar structures are promising candidates for advanced 3D random-access memory (RAM) architectures, offering high storage density, low-power consumption, and high operational speed. However, the width of uncharged Ising-Bloch domain walls in inorganic ferroelectrics (2 - 5 nm) is typically much smaller than that of domain walls and vortex cores in ferromagnets (~10-50 nm). This difference in characteristic length scales suggests that ferroelectric chiral polar structures could achieve significantly higher storage densities than those of ferromagnets. Because ultra-fast magnetic switching has been demonstrated both theoretically and experimentally [86, 87], ferromagnetic 3D-RAM may offer speed advantages over ferroelectric 3D-RAM [88, 89]. In contrast, it has been only recently that ultra-fast ferroelectric switching has been predicted [90]. However, multiaxial ferroelectric nanoparticles with low electric coercive fields could significantly improve the performance of a ferroelectric 3D-RAM, potentially bridging the performance gap.

Notably, vortex-type, labyrinth-type, and related chiral domain configurations offer potential for encoding information relevant to advanced cryptography [72 - 74]. Potential decoding can be performed by piezoresponse force microscopy (PFM) [91, 92], which provides information about polarization chirality and distribution with nanoscale resolution; and, complementary, by resonant elastic soft X-ray scattering [93, 94, 95], which is sensitive to chiral polar arrangements through dichroism effects.

It appears that quasi-continuum of multiple-degenerate labyrinth states may correspond to a negative dielectric susceptibility state stabilized by the presence of a screening shell [64]. The negative susceptibility can lead to a quasi-stationary negative capacitance state of the core-shell nanoparticles [78]. The state of negative differential capacitance (NC) was first predicted theoretically by Salahuddin and Datta [96] and much later realized experimentally in a bilayer capacitor made of paraelectric $SrTiO_3$ and ferroelectric $Pb_xZr_{1-x}TiO_3$ layers by Khan et al. [97, 98]. It was demonstrated experimentally that the total capacitance is greater than it would be for the $SrTiO_3$ layer of the same thickness as used in the bilayer capacitor, thus confirming the presence of a stable NC state in the Pb$_x$Zr$_{1-x}$TiO$_3$ layer, as evidenced by the increased overall capacitance. The realization of NC states has recently gained significant attention due to potential applications in advanced nanoelectronics. Replacing the standard insulator in a gate stack of a field-effect transistor (FET) with a ferroelectric NC insulator of the appropriate thickness offers several advantages. A key advantage of this approach is that it is a relatively simple replacement for conventional FETs; this substitution significantly reduces heat dissipation in nano-chips with a high density of critical electronic elements, and it reduces the FET subthreshold swing below the thermodynamics limit [99, 100].

While there are numerous experimental realizations of NC states in ferroelectric double-layer capacitors [101, 102, 103, 104], analytical descriptions remain limited. Due to the complexity of modeling domain structure emergence and stability, only a few studies have provided semi-analytical expressions for the onset of the NC effect and considered domain formation in the ferroelectric layer [105, 106, 107].

Analytical calculations based on the LGD free energy minimization corroborated by the FEM simulations show that ferroelectric BaTiO$_3$ nanocylinders, whose ends are covered by the thin paraelectric layer of SrTiO$_3$ (thickness $h_s \leq 10$ nm), can be suitable candidates for the controllable reduction of the SrTiO$_3$ layer capacitance due to the NC state emerging in the nanocylinders [78]. Short nanocylinders, whose length $h_c$ is smaller than its width $2R_c$ (see **Fig. 7(a)**), are preferable for the capacitor structures miniaturization. The LGD free energy potential of the NC states in BaTiO$_3$ nanocylinder has relatively flat negative wells, which appear due to the coupling of the positive parabolic potential of the SrTiO$_3$ layers with the double-well potential of the single-domain bulk BaTiO$_3$ (see red, blue, and green curves in **Fig. 7(b)**). As a result, the charge $Q$ stored on the electrodes covering the three-layer SrTiO$_3$-BaTiO$_3$-SrTiO$_3$ capacitor of the thickness $2h_s + h_c$ can exceed the charge $Q_r$ on the electrodes covering the SrTiO$_3$ layer of the thickness $2h_s$. The effective differential capacitance of an electroded system, $C_{eff}$, is equal to the first derivative of $Q$ with respect to the applied voltage $U$, $C_{eff} = \frac{dQ}{dU}$. If the voltage dependence $Q(U)$ is steeper than $Q_r(U)$, the differential capacitance of the SrTiO$_3$-BaTiO$_3$-SrTiO$_3$ capacitor can be larger than the capacitance $C_r = \frac{\varepsilon_0 \varepsilon_s}{2h_s}$ of the reference SrTiO$_3$ capacitor. The case $C_{eff} > C_r$ corresponds to the NC state, where the relative capacitance,

$$\Delta C = \frac{C_r - C_{eff}}{C_r}, \qquad (3)$$

is negative.

The physical origin of the NC state is the specific energy-degenerated metastable states of the spontaneous polarization in BaTiO$_3$ nanocylinders [78]. The metastable states appear dependent on the flexoelectric coupling and elastic strains induced by the presence of elastic defects in the shell. Elastic defects (e.g., dilatation centers, such as oxygen or cation vacancies, divacancies, OH-complexes, or isovalent impurity atoms) can create elastic strains in oxide ferroelectrics, which are usually called chemical (or compositional) strains [108, 109]. Choi et al. [110] experimentally demonstrated that with epitaxial stress (lattice mismatch) the spontaneous polarization $P_s$ of BaTiO$_3$ thin films, increased by 250 % due to epitaxial stress (lattice mismatch), while the Curie temperature $T_C$ rose to approximately 500°C. First principles calculations conducted by Ederer and Spaldin [111] along with phenomenological thermodynamic theory [112], corroborate the experimental results. In subsequent studies, it was demonstrated that the chemical strains induce a substantial increase of the Curie temperature (above 440 K) and tetragonality (up to 1.032) near the surface of a BaTiO$_3$ film with injected oxygen vacancies [113]. Furthermore, a lattice constants mismatch between the core and shell, which induces epitaxial strain, is responsible for the preservation and enhancement of $P_s$, tetragonality $c/a$ and $T_C$ increase in BaTiO$_3$ core-shell nanoparticles, as reported in Refs.[114, 115, 116]. The structural origin of recovered ferroelectricity and the phenomenological description of this striking effect in given by Zhang et al. [117] and Eliseev et al. [118], respectively.

Assuming that the formation energy of elastic defects is much smaller near the surface than in the bulk of the ferroelectric [119], elastic defects and corresponding chemical strains are accumulated in a thin shell under the surface of BaTiO$_3$ nanocylinders. Due to the electrostriction coupling and the flexoelectric effect, chemical strains can have a strong influence on the effective Curie temperature and the polarization distribution morphology and chirality [56, 80]. To emphasize the significance of this mechanism, we hereafter refer to it as "flexo-chemical" coupling [56].

As it was shown in Ref. [78], the nonzero components of the core strains, $u_i^c$, written in the Voigt notation, are:

$$u_3^c = u_c + (1 - \delta V)Q_{11}P_3^2 + \delta V \delta u, \qquad (4a)$$

$$u_1^c = u_2^c = (1 - \delta V)(u_c + Q_{12}P_3^2) + \delta V \left[u_s + \frac{(s_{11}-s_{12})\delta u + (s_{11}Q_{12}-s_{12}Q_{11})P_3^2}{2(s_{11}+s_{12})}\right]. \qquad (4b)$$

Here the relative shell volume ($\delta V$) and the "effective" strain ($\delta u$) are introduced as:

$$\delta V = \frac{V_s}{V}, \quad \delta u = u_s - u_c + u_m + u_t, \qquad (4c)$$

where the shell volume is $V_s$ and the core volume is $V$. The effective strain $\delta u$ is proportional to the difference between the core and the shell chemical strains ($u_c$ and $u_s$), as well as the contributions of the lattice constants mismatch ($u_m$) and/or different thermal expansion coefficients ($u_t$). Below we consider the simplest case where the elastic defects are postulated to be present only in the shell; other contributions to the effective strain $\delta u$ are absent, i.e., $\delta u \equiv u_s$ and $u_c = 0$ in Eqs.(4).

The FEM performed in Ref. [78] reveals that the effective strains $\delta u$ can induce vertex-like or vortex-like transitions of domain structure morphology in the BaTiO$_3$ core for zero or small flexoelectric coupling ($|F_{ij}| \leq 10^{-12} m^3/C$). Tensile strains induce and support the single-domain state in the central part of the core, while curled domain structures appear near the unscreened or partially screened ends of the BaTiO$_3$ nanocylinder. The vortex-like domains propagate toward the central part of the rod, completely filling the rod when it is covered with a compressed shell (see **Fig. 7(c)**). Under these conditions, the vortex intergrowth occurs for compressive chemical strains above some critical value (e.g., for $\delta u \leq -1\%$, as shown in **Fig. 7(c)**), which depends on the temperature, nanorod sizes, aspect ratio, and screening conditions at the nanorod ends (see details in Ref. [78]).

A relatively high flexoelectric coupling ($|F_{ij}| \geq 2 \cdot 10^{-11} m^3/C$) can lead to the appearance and stabilization of flexon-type [64] domain structure morphology in the BaTiO$_3$ nanocylinder core, examples are shown in **Fig. 7(d)** and **7(e)**. These polar textures reveal a chiral polarization structure of the axial polarization component $P_3$ containing two oppositely oriented diffuse axial domains located near the cylinder ends, separated by a region with a zero-axial polarization. The polarity of the axial domains can be switched by reversing the sign of the corresponding flexoelectric coefficient (see the inverted contrast in **Fig. 7(d)** and **7(e)**). Similar to the case shown in **Fig. 1**, flexons in the SrTiO$_3$-BaTiO$_3$-SrTiO$_3$ nanocapacitor display the polarization state of a vortex with an axially polarized core region located in the azimuthal plane, displaying characteristics of a meron structure. Moreover, the azimuthal distributions of the spontaneous polarization in-plane components $P_{1,2}$ do not significantly change in the presence of the flexoelectric coupling, and therefore they are not shown in **Fig. 7(d)** and **7(e).**

The dependence of the relative capacitance, $\Delta C$, on the effective strain $\delta u$ and thickness ratio $\frac{h_c}{h_s}$ calculated at room temperature is shown in **Fig. 7(f)**. A positive $\Delta C$ corresponds to a positive capacitance state, $C_{eff} < C_r$, which exists in the paraelectric (PE) phase of a bulk BaTiO$_3$ (see the lower rectangular region **Fig. 7(f)**). A negative $\Delta C$ corresponds to the NC state, $C_{eff} > C_r$, which exists between the black horizontal line and the black hyperbolic function. The hyperbolae denotes the boundary between the size-induced PE phase and the single-domain ferroelectric (FE) phase. Note that the dark-violet color in **Fig. 7(f)** corresponds to $\Delta C < -5$. The value of $C_{eff}$ sharply decreases in the dark-violet region and diverges ($\Delta C \to -\infty$) when approaching the PE-FE boundary. The white color corresponds to the region of the single-domain FE phase, where $\Delta C > 0$.

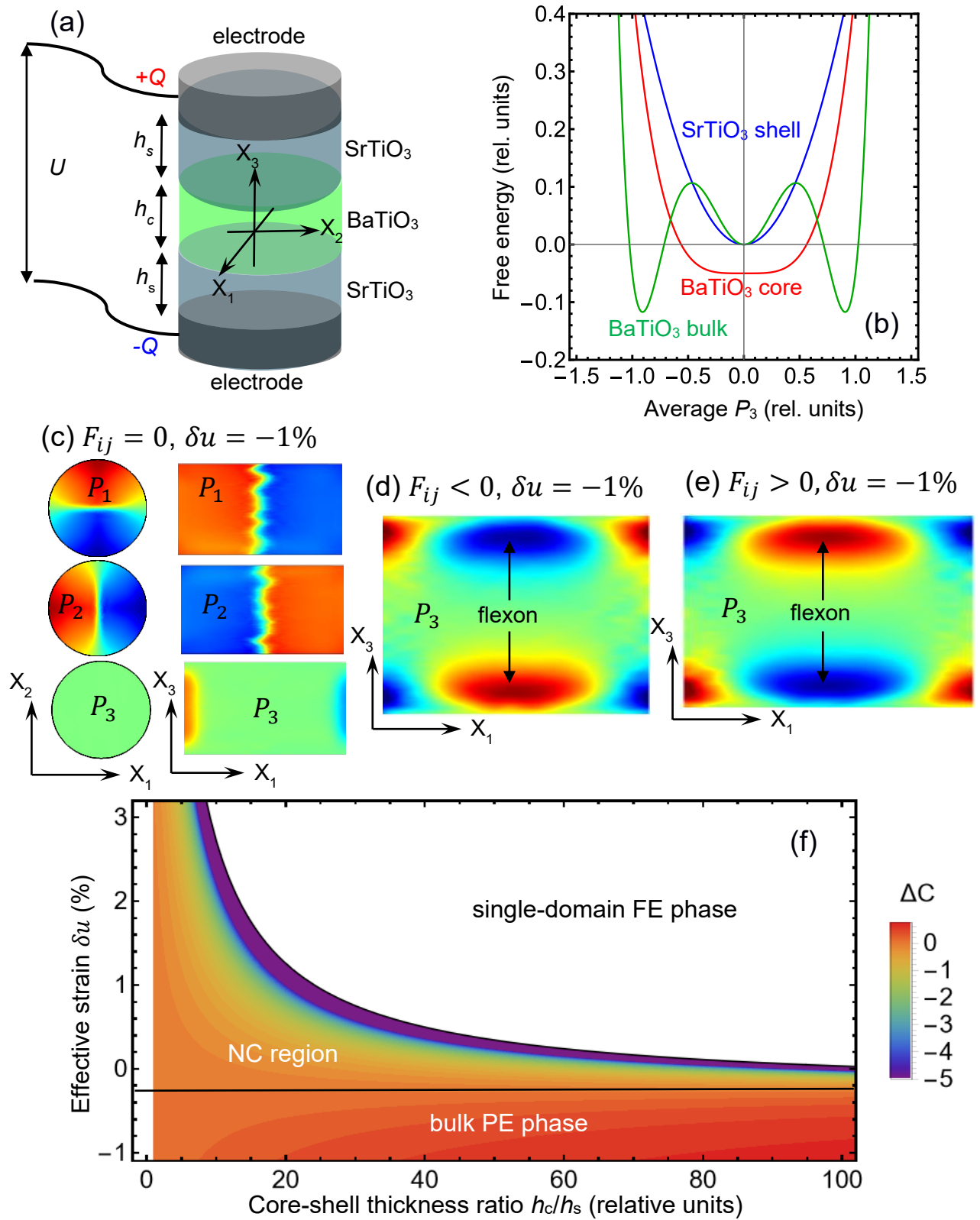

**FIGURE 7. The influence of flexo-chemical coupling on the polarization morphology and NC state in ferroelectric core-shell nanorods.** (a) Three-layer capacitor consisting of a BaTiO$_3$ nanocylinder, whose ends are covered by paraelectric SrTiO$_3$ layers. (b) Schematic illustration of the LGD free energy dependence on the polarization for single-domain bulk BaTiO$_3$ (green curve), paraelectric SrTiO$_3$ shell (blue curve), and BaTiO$_3$ core with metastable polarization states (red curve). (c) Typical distributions of spontaneous polarization components $P_i$ in the SrTiO$_3$-BaTiO$_3$-SrTiO$_3$ nanocapacitor calculated without the flexoelectric

coupling ($F_{ij} = 0$), but in the presence of compressive strain ($\delta u = -1\%$). Typical distributions of the axial spontaneous polarization $P_3$ in the nanocapacitor calculated for $\delta u = -1\%$, with **(d)** negative flexoelectric coefficients ($F_{11} = -3, F_{12} = -2.7, F_{44} = -9$ in $10^{-11} \text{m}^3/\text{C}$) and **(e)** positive flexoelectric coefficients ($F_{11} = 3, F_{12} = 2.7, F_{44} = 9$ in $10^{-11} \text{m}^3/\text{C}$). **(f)** The dependence of the relative capacitance, $\Delta C$, on the effective strain $\delta u$ and thickness ratio $\frac{h_c}{h_s}$ calculated for $T = 298$ K and relatively small flexoelectric coefficients $|F_{ij}| \leq 10^{-12} \text{m}^3/\text{C}$. The color scale illustrates the relative capacitance $\Delta C$. Parts (a)-(f) are adapted from Refs. [64] and [78].

It has been shown that thin films of Van der Waals ferrielectrics, such as a ferrielectric CuInP$_2$S$_6$ [120], are well-suitable for the analytical control of the NC effect [79] and subthreshold swing reduction in ferrielectric FETs [121], which has been observed experimentally in these layered materials [122, 123]. The theoretical work [79], which considers CuInP$_2$S$_6$ nanoflakes covered by an ionic charge with a surface charge density $\sigma_S$ (see **Fig. 8(a)**), reveals polar states with a relatively high polarization and stored free charge, mimicking "mid-gap" states associated with a surface field-induced transfer of Cu and/or In ions in the van der Waals gap. It is worth noting that a layered CuInP$_2$S$_6$ can be treated as a uniaxial ferrielectric with the spontaneous polarization directed normal to the layers, since its in-plane spontaneous polarization component is relatively small [120].

The Stephenson-Highland model [124] describes the relationship between surface charge density $\sigma_S[\delta\phi]$ and electric potential excess $\delta\phi$ at the nanoparticle surface. This model accounts for the coverages of positive ($i = 1$) and negative ($i = 2$) surface charges (e.g., ions) in a self-consistent manner. The corresponding Langmuir adsorption isotherm is given by [125, 126]:

$$\sigma_S[\delta\phi] \cong \sum_i \frac{eZ_i}{A_i}\left(1 + a_i^{-1} \exp\left[\frac{\Delta G_i + eZ_i \delta\phi}{k_B T}\right]\right)^{-1}, \qquad (5)$$

where $e$ is the electron charge, $Z_i$ is the ionization number of the adsorbed ions, $a_i$ is the dimensionless chemical activity of the ions in the environment (as a rule $0 \leq a_i \leq 1$)), $T$ is the absolute temperature, $A_i$ is the area per surface site for the adsorbed ion, and $\Delta G_i$ are the formation energies of the surface charges (e.g., ions and/or electrons) at normal conditions; the subscript $i = 1, 2$. Mismatch and/or chemical strains exist at the nanoflake-substrate interface. From Eq.(5), the effective screening length $\lambda$ associated with the adsorbed charges is $\frac{1}{\lambda} \approx \sum_i \frac{(eZ_i)^2 a_i \exp\left[\frac{\Delta G_i}{k_B T}\right]}{\varepsilon_0 k_B T A_i \left(a_i + \exp\left[\frac{\Delta G_i}{k_B T}\right]\right)^2}$ [126]. The tunable surface chemical activity enables the use of the flexo-sensitive ferroelectric nanoparticles for catalysis [127].

Consideration of the flexoelectric coupling leads to the appearance and stabilization of tubular-like domain patterns in the CuInP$_2$S$_6$ nanoflake, examples are shown in **Fig. 8(b)** and **8(d)**. The polarity of the domains can be switched by reversing the sign of the corresponding flexoelectric coefficient (see the inverted contrast in **Fig. 8(b)** and **8(d)**). Without flexoelectric coupling, the domains have an irregular shape (see **Fig. 8(c)**). The flexoelectric coupling controls the domain structure polarity due to the flexoelectric field $E_l^{flexo}$, which is proportional to the convolution of the flexoelectric tensor coefficients $F_{ijkl}$ with the gradient of electric stress $\frac{\partial \sigma_{ij}}{\partial x_k}$, namely $E_l^{flexo} \sim F_{ijkl} \frac{\partial \sigma_{ij}}{\partial x_k}$. Lattice mismatch and/or chemical strains $u_{ij}$, which are concentrated at the nanoflake-substrate interface, are the source for the flexoelectric field in the considered case. The statement is illustrated by the distribution of strain components $u_{ij}$ in axial and equatorial cross-sections of the CuInP$_2$S$_6$ nanoflake, shown in **Fig. 9**. It is seen from **Fig. 9(d)-(i)** that the strain components and their gradients reach maximal values near the nanoflake-substrate interface and minimal values at the domain walls. Following Hooke's law, the stresses $\sigma_{ij}$ are proportional to the strains $u_{kl}$, namely $\sigma_{ij} \sim c_{ijkl} u_{kl}$, where $c_{ijkl}$ are components of the elastic stiffness tensor. Thus, the flexoelectric field is maximal near the nanoflake-substrate interface, where the mismatch and/or chemical strain are applied.

Changes in the magnitude of ionic screening, lattice mismatch, and/or chemical strains in CuInP$_2$S$_6$ nanoflakes induce a broad range of transitions. These transitions occur between paraelectric (PE), antiferroelectric (AFE), ferrielectric (FI1 and FI2) phases, and ferri-ionic (FII1 and FII2) states [79]. The existence and diversity of these states determine the controllable NC effect of CuInP$_2$S$_6$ nanoflakes. Furthermore, a dielectric layer and an ionic-electronic charge layer at the CuInP$_2$S$_6$ surface can induce a "shallow" PE-like state characterized by a plateau-like potential well located near zero energy.

A typical example of the plateau-like free energy potential well, calculated for an averaged value of $P_3$, is shown by the red curve in **Fig. 8(e)**. The plateau-like potential well appears due to the coupling between the CuInP$_2$S$_6$ potential with the four negative wells (shown by the green curve in **Fig. 8(e)**) and the positive parabolic potential of the dielectric layer (DL) (shown by the blue curve in **Fig. 8(e))**. The energy plots shown in **Fig. 8(e)** are qualitatively similar to those of **Fig. 7(b)**. The quantitative differences between the green and red curves in **Fig. 8(e)** and **Fig. 7(b)** are related to the presence of ionic surface charge (SC), which plays an important role in making the total potential relief of the bi-layer system significantly flatter for certain parameters of surface charge and layer thicknesses (see details in Ref. [79]).

As a result of the potential well flattering, the free charge $q = \frac{q_2 - q_1}{2}$ stored at the electrodes covering the bi-layer system "CIPS+SC+DL" becomes larger than the "reference" charge at the electrodes covering the dielectric layer. The NC effect occurs when the value of $C_{eff}$ exceeds the capacitance $C_r$ of the reference dielectric capacitor of thickness $d$. The dependence of the relative capacitance, $\Delta C$, on the chemical activity of the ions $a_i$ and the strain $\delta u$ is shown in **Fig. 8(f)**. CuInP$_2$S$_6$ material parameters, sizes, and surface charge characteristics are listed in Ref. [79]. The NC region, where $\Delta C < 0$, is located inside the triangular-like region with the black boundary. The PE-like state with the flat well, where the NC effect is pronounced ($\Delta C < -1$), is shown by the dark-violet color in **Fig. 8(f)**. The strain range of the NC state stability is widest for $a_i < 10^{-2}$ (namely $0 \leq \delta u \leq 0.55$ %), its width strongly decreases as $a_i$ increases above $10^{-1}$; the NC state gradually disappears for $a_i \to 1$. As shown in Ref. [79], the NC effect can exist over a wide range of dielectric layer and nanoflake thicknesses, strains, and surface charge parameters.

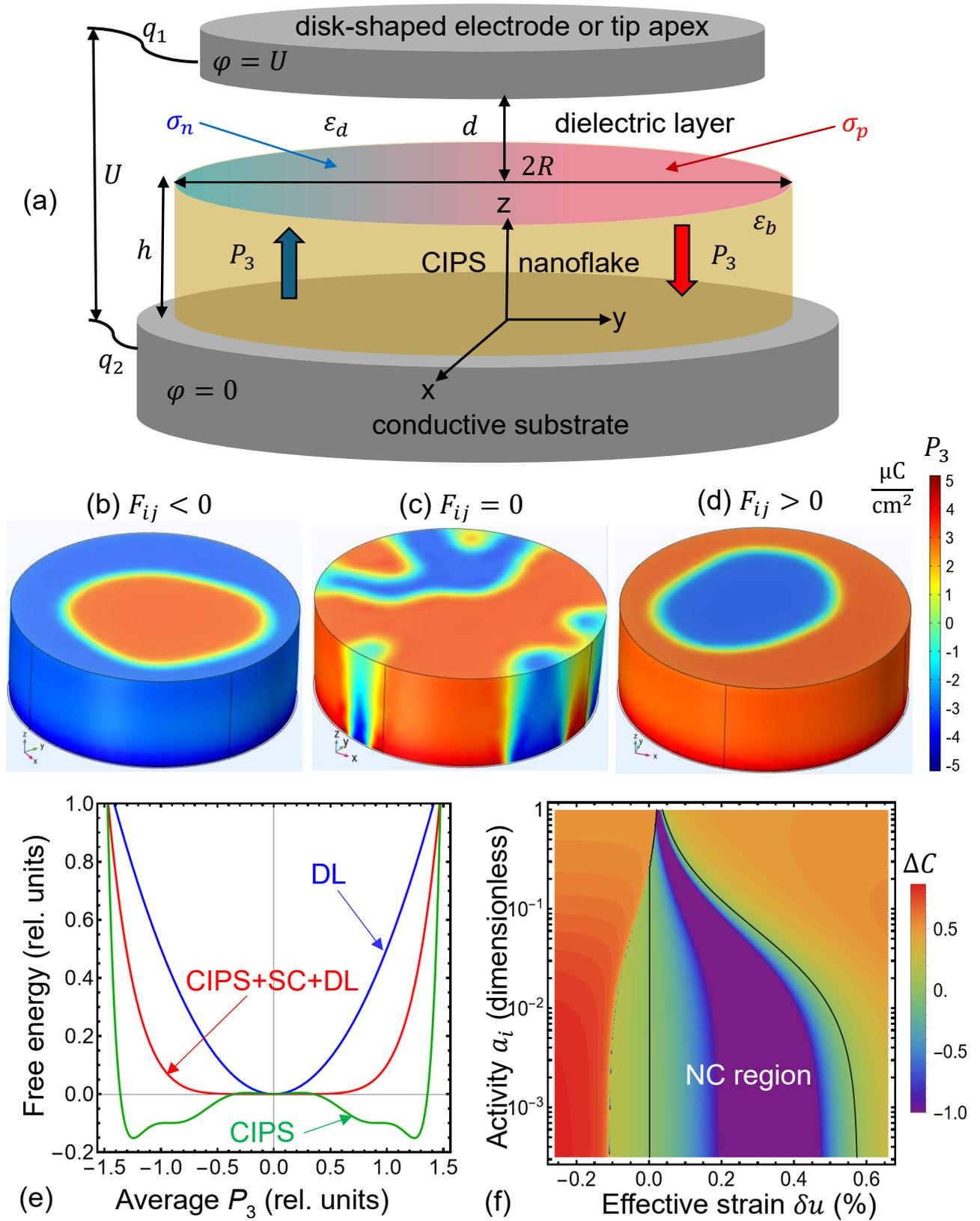

**FIGURE 8**. **The influence of flexo-chemical coupling on the polarization morphology and NC state in CuInP$_2$S$_6$ nanoflakes.** **(a)** A cylindrical-shaped CuInP$_2$S$_6$ (CIPS) nanoflake of height $h$ and radius $R$ placed between a top electrode, dielectric layer of thickness $d$, and a conducting substrate. The "up" and "down" directions of the uniaxial polarization $P_3$ are shown by the blue and red arrows, respectively. The surface of the nanoflake is covered with a layer of mobile surface charges, whose density $\sigma_S$ is the sum of positive (red

color) and negative (blue color) charges, $\sigma_p$ and $\sigma_n$, respectively. Typical distributions of the spontaneous polarization component $P_3$ in the CIPS nanoflake calculated for **(b)** negative flexoelectric coefficients ($F_{13} = -1.61, F_{23} = -1.78, F_{33} = 14.4, F_{44} = -14.3, F_{55} = -14.9$ in $10^{-11}\text{m}^3/\text{C}$), **(c)** zero ($F_{ij} = 0$), and **(d)** positive flexoelectric coefficients ($F_{13} = 1.61, F_{23} = 1.78, F_{33} = 14.4, F_{44} = 14.3, F_{55} = 14.9$ in $10^{-11}\text{m}^3/\text{C}$). Effective strain $\delta u = 1\%$, screening length $\lambda = 3$ pm, nanoflake height $h = 7$ nm, radius $R = 10$ nm, dielectric layer thickness $d = 1$ nm and temperature $T = 298$ K. **(e)** Schematic illustration of the free energy dependence on the average polarization for single-domain bulk CuInP$_2$S$_6$ with two stable and two metastable polarization states (green curve marked as "CIPS"), the dielectric layer (blue curve marked as "DL"), and the bi-layer structure consisting of the CIPS nanoflake covered by surface charge and a dielectric layer (red curve marked as "DL+SC+CIPS"). **(f)** The dependence of the relative capacitance, $\Delta C$, on the surface charge activity $a$ and effective strain $\delta u$. The sizes: $h = 10$ nm, $d = 1$ nm, and $R = 0.5$ μm. Parts (a), (e), and (f) are adapted from Ref. [79].

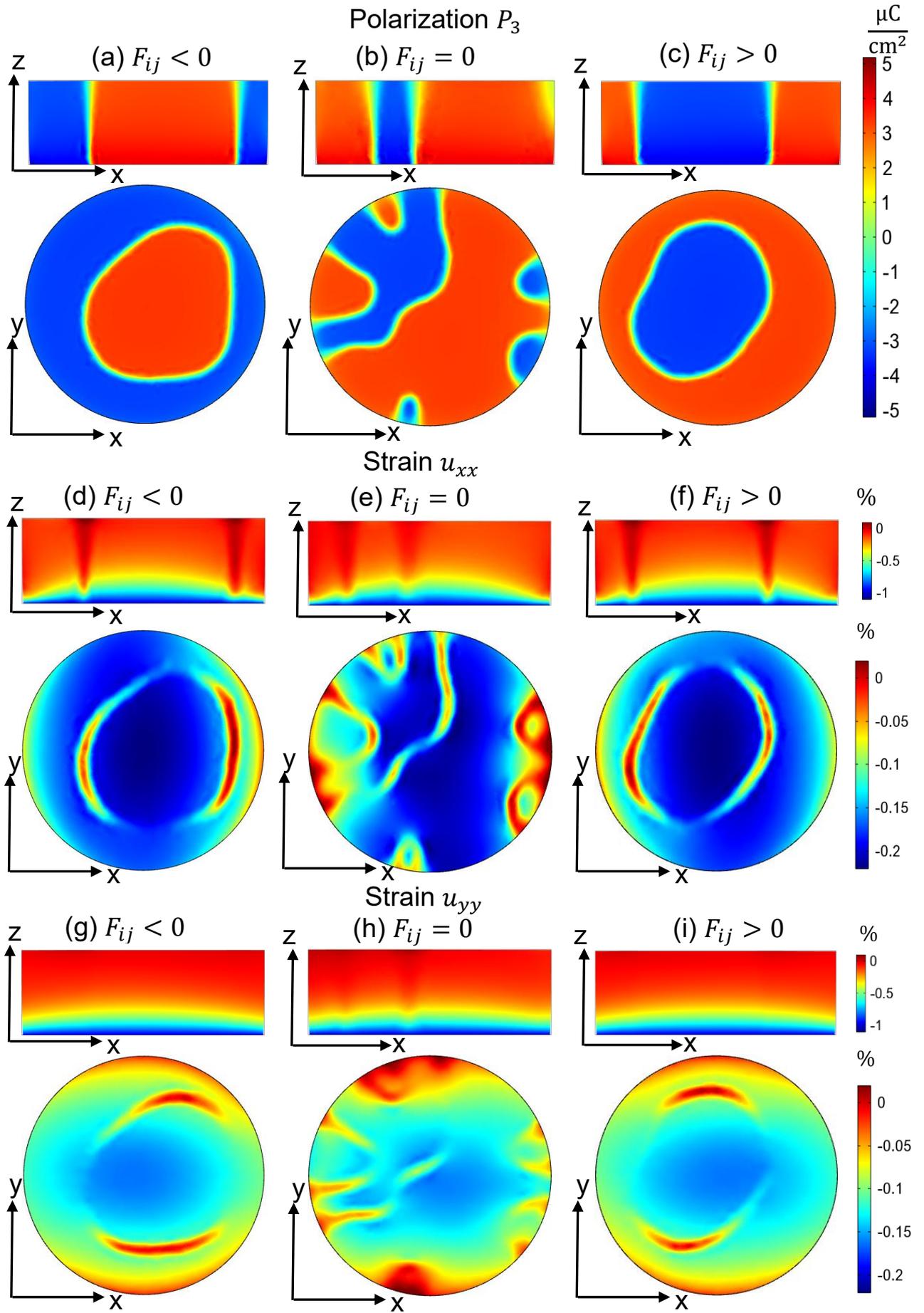

**FIGURE 9**. Typical distributions of the spontaneous polarization component $P_3$ (**a, b, c**), strain components

$u_{xx}$ (**d, e, f**), and $u_{yy}$ (**g, h, i**) in the axial ZX and equatorial XY cross-sections of the CuInP$_2$S$_6$ nanoflake calculated for negative **(a, d, g)**, zero **(b, e, h)**, and positive **(c, f, i)** flexoelectric coupling coefficients $F_{ij}$. Parameters are the same in **Fig. 8**.

### 3. Conclusions

To summarize, the flexoelectric effect significantly affects the polarization structure in thin films and core-shell nanoparticles of multiaxial ferroelectrics. The flexoelectric coupling leads to a noticeable curling of the flux-closure domain walls in multiaxial ferroelectric thin films and significantly impacts the manifold degenerated three-dimensional vortex states in corresponding core-shell nanoparticles. Under certain conditions, flexo-sensitive polarization vortices can form in multiaxial ferroelectrics. Furthermore, an anisotropic flexoelectric effect can give rise to meron-type polarization states with distinct chiral properties in multiaxial ferroelectrics, termed flexons.

In core-shell nanoparticles of uniaxial ferroelectrics, flexoelectric coupling can control the polarity and chirality of equilibrium polar domain structures. Specifically, we observe a gradual transition of labyrinthine domains from a branched-like to a spiral-like morphology as the flexo-coupling strength increases. The transition of the domain structure morphology occurs in relatively large nanoparticles but disappears as the core radius decreases. This disappearance is due to the increase of the effective coefficients in polarization gradient energy, which are renormalized proportionally to the second power of the flexoelectric coefficients. Consequently, the transition is virtually insensitive to the sign of the flexoelectric coefficients.

The combined action of the flexoelectric effect, electrostriction coupling, lattice mismatch [128] and chemical strains, termed as flexo-chemical coupling, significantly influence the effective Curie temperature, polarization distribution, domain morphology, and chirality in core-shell ferroelectric nanoparticles. In particular, flexo-chemical coupling leads to the appearance and stabilization of chiral polarization morphology in van der Waals ferrielectrics nanoflakes covered with a shell of ionic-electronic screening charge. Varying the magnitude of ionic screening, lattice mismatch, and chemical strains induces a broad range of transitions between paraelectric, antiferroelectric and ferrielectric phases, and ferri-ionic states in the nanoflakes.

Applications of the flexo-sensitive chiral polar structures in core-shell ferroelectric nanoparticles hold significant promise for nanoelectronics and catalysis. For instance, it appeared that multiple-degenerated labyrinthine states may correspond to the NC state stabilized in the uniaxial ferroelectric core by the presence of a screening shell. Also, we expect that the flexoelectric control of labyrinthine domains is possible in many types of core-shell ferroelectric nanoparticles, suggesting potential applications in advanced cryptography. The paraelectric-like state of van der Waals ferrielectric nanoflakes covered by shells of ionic-electronic screening charge reveals a pronounced NC effect existing over a relatively wide range of nanoflake thicknesses, mismatch and chemical strains, and surface charge densities.

**Acknowledgements.** A.N.M. acknowledges EOARD project 9IOE063b and related STCU partner project P751b, and the National Research Foundation of Ukraine (project "Manyfold-degenerated metastable states of spontaneous polarization in nanoferroics: theory, experiment, and perspectives for digital nanoelectronics," grant N 2023.03/0132). A.N.M. also acknowledges the Horizon Europe Framework Programme (HORIZON-TMA-MSCA-SE), project № 101131229, Piezoelectricity in 2D-materials: materials, modeling, and applications (PIEZO-2D). The work of E.A.E. is funded by the National Research Foundation of Ukraine (project "Silicon-compatible ferroelectric nanocomposites for electronics and sensors", grant N 2023.03/0127). R.H. acknowledges support by the French National Research Agency (ANR) under Contract No. ANR-24-CE30-5392 through the TopoTherm project.

[128] Note that if the system reaction to the mismatch strain is the appearance of homogeneous strains (such as in thin films without dislocations, domain structure, and top stress-free surface), "flexo-mismatch" strains do not emerge, because the flexo-coupling requires a strain gradient or polarization gradient. However, if a lattice mismatch leads to the inhomogeneous strains, they become coupled with the flexoelectricity.